\documentclass[11pt,reqno,a4paper]{amsart}

\usepackage[margin=1.1in]{geometry}

\usepackage{graphicx}
\usepackage{multicol,multirow}
\usepackage{amsmath,amssymb,amsfonts}
\usepackage{mathrsfs}
\usepackage{amsthm}
\usepackage{rotating}
\usepackage{appendix}
\usepackage[numbers,super,sort&compress,round]{natbib}
\usepackage{ifpdf}
\usepackage[T1]{fontenc}
\usepackage{times}
\usepackage{newtxmath}
\usepackage{textcomp}
\usepackage[table]{xcolor}
\usepackage{hyperref}
\usepackage{cleveref}
\usepackage{lipsum}
\usepackage[ruled,linesnumbered, noline]{algorithm2e}
\Crefname{algocf}{Algorithm}{Algorithms}

\begin{document}

\title{\textbf{Moment-based metrics for molecules computable from cryo-EM images}}

\author[A. Zhang]{Andy Zhang}
\address{Program in Applied and Computational Mathematics, Princeton University, Princeton, 08544, NJ, USA}
\email{az8940@princeton.edu}
\author[O. Mickelin]{Oscar Mickelin}
\address{Program in Applied and Computational Mathematics, Princeton University, Princeton, 08544, NJ, USA}
\email{hm6655@princeton.edu}
\author[J. Kileel]{Joe Kileel}
\address{Department of Mathematics and Oden Institute, University of Texas at Austin, Austin, 78712, TX, USA}
\email{jkileel@math.utexas.edu}
\author[E. J. Verbeke]{Eric J. Verbeke}
\address{Program in Applied and Computational Mathematics, Princeton University, Princeton, 08544, NJ, USA}
\email{ev9102@princeton.edu}
\author[N. F. Marshall]{Nicholas F. Marshall}
\address{Department of Mathematics, Oregon State University, Corvallis, 97331, OR, USA}
\email{marsnich@oregonstate.edu}
\author[M. A. Gilles]{Marc Aur{\`e}le Gilles}
\address{Program in Applied and Computational Mathematics, Princeton University, Princeton, 08544, NJ, USA}
\email{mg6942@princeton.edu}
\author[A. Singer]{Amit Singer}
\address{Department of Mathematics and Program in Applied and Computational Mathematics, Princeton University, Princeton, 08544, NJ, USA}
\email{amits@math.princeton.edu}

\begin{abstract}
Single particle cryogenic electron microscopy (cryo-EM) is an imaging technique capable of recovering the high-resolution 3-D structure of biological macromolecules from many noisy and randomly oriented projection images. One notable approach to 3-D reconstruction, known as Kam’s method, relies on the moments of the 2-D images. Inspired by Kam's method, we introduce a rotationally invariant metric between two molecular structures, which does not require 3-D alignment.  Further, we introduce a
metric between a stack of projection images and a molecular structure, which is invariant to rotations and reflections and does not require performing 3-D reconstruction.
Additionally, the latter metric does not assume a uniform distribution of viewing angles.
We demonstrate uses of the new metrics on synthetic and experimental datasets,  highlighting their ability to measure structural similarity. 
\end{abstract}


\maketitle

\section{Introduction}
Single particle cryogenic electron microscopy (cryo-EM) enables high-resolution reconstruction of 3-D biological macromolecules from a large collection of noisy and randomly oriented projection images. Kam’s method \cite{Kam} is one of the earliest methods proposed for homogeneous reconstruction in cryo-EM. It is a statistical method-of-moments approach applied to the cryo-EM reconstruction problem, where the computation of low-order statistics of 2-D images allows for the estimation of 3-D structure through solving a polynomial system. Kam’s method has helped push the theoretical understanding of the reconstruction process - under certain conditions, it is a provable algorithm and provides bounds for the estimated structure's quality in terms of the noise level and the number of images \cite{bhamre2017anisotropic,bhamre2015orthogonal,levin20183d, bandeira2023estimation, huang2023orthogonal,bendory2023autocorrelation, nonuniform}. Kam's method also enjoys other remarkable properties:
\begin{enumerate}
\item It bypasses the need for angular assignment, typically a large computational burden in competing methods.
\item It is a streaming algorithm and is thus theoretically much faster than iterative methods.
\item It can – in theory – break the detection limit of the minimal size of proteins that can be observed in cryo-EM \cite{bendory_siam}.
\end{enumerate}
While theoretically attractive, Kam's method has not been able to yield high-resolution reconstructions as yet. One direction that is currently being explored to resolve this issue is the development of better priors, for instance, based on the sparsity of the solution as in \cite{bendory2023autocorrelation}. Separately, there has been considerable, continued interest in utilizing the vast amount of solved structures stored in the Protein Data Bank (PDB)~\cite{berman_protein_2000} to improve cryo-EM reconstructions.

Leveraging the PDB as a prior, we propose a method to match either projection images or molecular volumes to a database of previously solved structures (Section~\ref{sec:def}). 
We use this procedure as a rotationally and reflectionally invariant metric that can be directly computed from raw image datasets without needing a 3-D reconstruction process. 
Importantly, our metric neither relies on prior knowledge of rotations nor assumes a uniform rotational distribution, making it applicable to typical datasets.

To demonstrate the efficacy of our metric, we compare it to existing methods and show empirically that it achieves similar performance to alignment-based metrics.
As an application, we use our metric to compute a low-dimensional embedding of a subset of the PDB into the Euclidean plane, visually showcasing how structurally similar proteins are embedded near each other (Section~\ref{sec:database}). 
Further, we apply the version of the metric that can be directly computed from stacks of 2-D images, and show that it gives an efficient methodology to identify the nearest neighbors in a database to a given set of experimental moments on synthetic and real datasets (Sections~\ref{sec:image-metric} and \ref{sec:experimental}).

\section{Background}
This section presents the mathematical preliminaries needed to define our metric.
Let $\Phi : \mathbb{R}^3 \rightarrow \mathbb{R}$ be the 
electrostatic potential of a molecule and 
 $\widehat{\Phi} : \mathbb{R}^3 \rightarrow \mathbb{C}$ be its Fourier transform, which we define by
$$
\widehat{\Phi}(\xi) = \int_{\mathbb{R}^3} \Phi(x) e^{-i \xi \cdot x} dx.
$$
A single projection image is given by 
$$I = h \ast P R\Phi + \varepsilon,$$
and its Fourier transform is
$$ \widehat{I} = H \cdot SR \widehat{\Phi} + \widehat{\varepsilon},$$
where $P$ is the projection operator, $S$ is the slicing operator, $h$ is a point spread function, $H$ is the contrast transfer function (CTF), $\varepsilon$ is noise, and $R \in \operatorname{SO}(3)$ is a rotation operator.
We assume that the Fourier transform $\widehat{\Phi}$ can be expanded in a spherical harmonics expansion:
\begin{equation} \label{eq:hatPhi}
\widehat{\Phi}(r, \theta, \varphi) = \sum_{\ell=0}^L\sum_{m = -\ell}^{\ell} A_{\ell,m}(r) Y_\ell^m(\theta, \varphi),
\end{equation}
where $(r,\theta,\phi)$ are spherical coordinates, and $Y_\ell^m$ denotes the complex spherical harmonic:
$$
Y_\ell^m(\theta,\varphi) := \left( \frac{(\ell -m)! (2 \ell +1)}{4\pi (\ell+m)!} \right)^{1/2} e^{i m \varphi} P_\ell^m (\cos \theta),
$$
where $P_\ell^m$ are the associated Legendre polynomials,
$A_{\ell,m}(r)$ are $r$ dependent coefficients, and $L$ is a bandlimit parameter. See Eq.~14.30.1 in \cite{DLMF} for the definitions of $Y_\ell^m$ and $P_\ell^m$.

Let $\rho : \operatorname{SO}(3) \rightarrow \mathbb{R}$ be the probability density function of the rotational distribution, which without loss of generality is invariant to in-plane rotations and reflections.  (Note that by augmenting the dataset with in-plane rotations and reflections of all 2-D images, one can always reduce to the case of such an invariant distribution $\rho$, e.g. see \cite{ponce2011computing}.) More formally, $\rho$ is a function on $SO(3)/O(2) \simeq \mathbb{S}^2 / \{ \pm 1 \}$ which is formed by identifying antipodal points on the sphere $\mathbb{S}^2$ \cite{Bredon1993}. Thus, we model the density as a function $\rho: \operatorname{SO}(3) \rightarrow \mathbb{R}$ with an even-degree spherical harmonics expansion:
\begin{equation}\label{eq:rho}
\rho(R) = \sum_{\ell=0}^P \sum_{m=-2\ell}^{2\ell} B_{2\ell,m} \overline{Y_{2\ell}^m}(\theta(R), \varphi(R)),
\end{equation}
where $(\theta(R), \varphi(R))$ represent the third column of the rotation matrix given by $R$ in spherical coordinates, and $P \in \mathbb{Z}_{\geq 0}$ is a bandlimit parameter (see Section 4.2 in \cite{nonuniform}). The analytical first and second moments $\mathbf{m}_1$ and $\mathbf{m}_2$ of the Fourier-transformed projection images with respect to $\rho$ are 
\begin{equation}
\label{eq:moments}
\begin{split}
\mathbf{m}_1(x,y) &= \int (R \cdot \widehat{\Phi})(x,y,0) \rho(R) dR,  \qquad \text{and} \\
\mathbf{m}_2\left((x,y),(x',y') \right) &= \int (R \cdot\widehat{\Phi})(x,y,0) (R \cdot \widehat{\Phi})(x',y',0) \rho(R) dR,
\end{split}
\end{equation}
where $dR$ denotes integration with respect to the Haar measure on $SO(3)$. It will be convenient to write $(x,y)$ and $(x',y')$ in terms of polar coordinates $(r,\phi)$ and $(r',\phi')$, respectively. 
In Appendix~\ref{app:moments}, we show in Eq.~\eqref{eq:first-moment} and~\eqref{eq:second-moment} that the first moment only  depends on $r$, i.e., $\mathbf{m}_1 = \mathbf{m}_1(r): \mathbb{R}_{\geq 0} \rightarrow \mathbb{C}$, and that the second moment only depends on $r,r'$ and $\Delta\phi = \phi - \phi'$, i.e., $\mathbf{m}_2 = \mathbf{m}_2(r, r',\Delta\phi): \mathbb{R}_{\geq 0} \times \mathbb{R}_{\geq 0} \times [-2\pi, 2\pi] \rightarrow \mathbb{C}$. We write $\mathbf{m}_1 = \mathbf{m}_1(\widehat{\Phi},\rho)$ 
and $\mathbf{m}_2 = \mathbf{m}_2(\widehat{\Phi},\rho)$ to denote the first and second moments defined by $\widehat{\Phi}$ and $\rho$ 
when discussing multiple structures.
The basis of Kam's method is that the moments in Eq.~\eqref{eq:moments} can be estimated from images and related to expansion coefficients for the potential $\widehat{\Phi}$, see \Cref{sec:methodology} for explicit formulas.

\section{Definition of Kam's metrics}  \label{sec:def}
We now use metrics between the moments in Eq.~\eqref{eq:moments} to define similarity between proteins as well as stacks of images.
A first function is used to measure similarity of two known structures by the moments of their potential as defined in Eq.~\eqref{eq:moments}. The second is used to measure similarity between a known structure and the unknown structure observed in a dataset of images.

Crucially, the metrics can be computed without performing 3-D alignment of the structures, reducing their computational costs compared to other approaches. Moreover, one of the metrics can be directly computed from noisy and CTF-affected projection images. This enables a nearest neighbor search among known structures to determine an initialization for the 3-D reconstruction pipeline, especially in the expectation maximization procedure \cite{scheres2012relion, scheres2012bayesian}.

\subsection{Kam's volume metric $d_{\operatorname{vKam}}$}\label{sec:volume-metric}
Here we introduce the first of Kam's metrics, which measures the similarity of two 3-D structures. We use this to perform dimensionality reduction to visualize the relation between structures from a subset of the PDB.

In detail, given two 3-D structures $\Phi_1$ and $\Phi_2$, we define the distance between them through their first and second moments $\mathbf{m}_1$ and $\mathbf{m}_2$ under a uniform distribution of viewing directions which we denote by $\rho = \rho_u$. We will derive the explicit equations for the uniform case in Eq.~\eqref{eq:first-moment-unif} and \eqref{eq:second-moment-unif}. We then measure the resulting weighted deviation of the first and second moments by
\begin{align} \label{eq:continuous_vKam} \| \mathbf{m}_2(\widehat{\Phi}_1, \rho_u) - \mathbf{m}_2(\widehat{\Phi}_2, \rho_u) \|_{L^2(\mathbb{R}^2 \times \mathbb{R}^2)}^2  + \lambda  \| \mathbf{m}_1(\widehat{\Phi}_1, \rho_u) - \mathbf{m}_1(\widehat{\Phi}_2, \rho_u) \|_{L^2(\mathbb{R}^2)}^2,
\end{align} 
where $\lambda \geq 0$ is a hyperparameter which we set to 1 for all experiments.
The moments will be represented on a discretized voxel grid, and we therefore replace the continuous norms with discrete norms. More specifically, we will represent the second moment using a grid $r, r' \in \{r_1, \ldots , r_{\lfloor N/2 \rfloor} \}$ and $\Delta \phi := \phi - \phi' \in \{\Delta \phi_1 , \ldots , \Delta \phi_N\}$, where $N$ is the number of pixels of one side of the discretized volume. We define the grid points $r_k = \delta k / N$, $\Delta\phi_j = 2\pi j / N - 2\pi$, for $k = 1, \ldots , \lfloor N/2 \rfloor$, and $j = 1, \ldots , 2N$, where $\delta$ is the side length of the volume grid in angstroms. We then use the following two approximations to the continuous norms above
\begin{equation} \label{eq: discrete_vKam}
 \| \mathbf{M} \|_{w_2}^2 := \sum_{\substack{j = 1, \ldots, 2N \\ k_1 = 1, \ldots, \lfloor N/2 \rfloor \\ k_2 = 1, \ldots, \lfloor N/2 \rfloor}} \left|\mathbf{M}(\Delta \phi_j,r_{k_1},r_{k_2})\right|^2 r_{k_1} r_{k_2}, \quad  \| \mathbf{N} \|_{w_1}^2 := \sum_{k = 1, \ldots, \lfloor N/2 \rfloor} \left|\mathbf{N}(r_{k})\right|^2 r_{k} .
\end{equation}
With these norms, we define the metric comparing two sets of moments of two 3-D structures by
\begin{align}
\label{eq:d-Kam1}
d_{\operatorname{vKam}} ( {\Phi}_1, {\Phi}_2 )  :=  \left( \| \mathbf{m}_2(\widehat{\Phi}_1, \rho_u) - \mathbf{m}_2(\widehat{\Phi}_2, \rho_u) \|_{w_2}^2  + \lambda  \| \mathbf{m}_1(\widehat{\Phi}_1, \rho_u) - \mathbf{m}_1(\widehat{\Phi}_2, \rho_u) \|_{w_1}^2 \right)^{1/2},
\end{align}

This distance is rotationally invariant since for any rotation $R$, we have $\widehat{R\cdot \Phi} = R\cdot \widehat{\Phi}$ and the moments $\mathbf{m}_1$ and $\mathbf{m}_2$ in Eq.~\eqref{eq:rho} satisfy 
\begin{equation}\label{eq:rotate_moments}
    \mathbf{m}_i(\widehat{R\cdot \Phi}, \rho) = \mathbf{m}_i(\widehat{ \Phi}, R^T\cdot \rho),
    \end{equation}
    as can be seen through a change of variables in Eq.~\eqref{eq:moments}. When $\rho = \rho_u$ is uniform, clearly $R^T\cdot \rho  = \rho$, which therefore shows rotational invariance of the cost function in Eq.~\eqref{eq:continuous_vKam}, up to the discretization of the volume grid. Note that this bypasses the need for an alignment step. We detail the procedure for computing $\mathbf{m}_1$, $\mathbf{m}_2$ and therefore $d_{\operatorname{vKam}}$ in ~\Cref{app:moments}. 
Under certain conditions, it has been demonstrated that the second moment of the image collection identifies the 3-D structure uniquely \cite{bhamre2017anisotropic,bhamre2015orthogonal,levin20183d, huang2023orthogonal,bendory2023autocorrelation} or up to a finite list of candidate structures \cite{nonuniform}. In section \Cref{sec:database}, we show that our metric is alike other similarity scores but remarkably doesn't rely on alignment.

\subsection{Kam's image metric $d_{\operatorname{iKam}}$}

We now introduce a metric between the empirical moments computed from a set of experimental projection images to the moments computed from the atomic coordinates of a known structure that compares images to the known structure.  We detail the procedure for computing these moments in ~\Cref{app:moments}.

 If the distribution of poses in the experimental dataset would be known to be uniform, the empirical moments could directly be substituted for $\mathbf{m}_1$ and $\mathbf{m}_2$ in Eq.~\eqref{eq:d-Kam1} and the pseudometric could be defined as the deviation between the moments of the two structures. In practice, however, the distribution of angles is not uniform and is unknown. Since the moments are functions of this distribution, it must therefore be inferred. 
 
We will show in Eq.~\eqref{eq:first-moment} and Eq.~\eqref{eq:second-moment} that $\mathbf{m}_1$ and $\mathbf{m}_2$ depend linearly on the expansion coefficients $B_{p,u}$ of the distribution of viewing directions. The optimization problem minimizing the discrepancy between the moments of the two structures is, therefore, a linear least-squares problem in $B_{p,u}$. It follows from Table~3 of \cite{nonuniform}, that this linear least squares is Zariski-generically full-rank (although not necessarily well-conditioned) for various small bandlimits $L$ and $P$. 
Solving this optimization problem efficiently eliminates the unknown rotational distribution. We then define the metric between the moments of the structure $\Phi$ and the experimental moments $\tilde{\mathbf{m}}_1, \tilde{\mathbf{m}}_2$ by

\begin{align} \label{eq:continuous_iKam}
\min_{\rho \in \mathcal{P}}  \| \tilde{\mathbf{m}}_2 - \mathbf{m}_2 (\widehat{\Phi}, \rho) \|_{L^2(\mathbb{R}^2 \times \mathbb{R}^2)}^2  +  \lambda  \| \tilde{\mathbf{m}}_1 - \mathbf{m}_1(\widehat{\Phi}, \rho) \|_{L^2(\mathbb{R}^2)}^2,
\end{align}
where $\lambda \geq 0$ is a hyperparameter which we set to 1 for all experiments and
\begin{equation}\label{eq:admissible}
    \mathcal{P} = \left\{\rho(R): \operatorname{SO}(3) \rightarrow \mathbb{R} \text{ s.t. } \rho(R) = \sum_{\ell=0}^P \sum_{m=-2\ell}^{2\ell} B_{2\ell,m} \overline{Y_{2\ell}^m}(\theta(R),\varphi(R)),  \text{ and }  \int_{\operatorname{SO}(3)} \!\!\!\!\rho(R) dR = 1 \right\},
\end{equation} is the set of admissible distributions of viewing directions that are invariant to global reflections and in-plane rotations, where $(\theta(R), \varphi(R))$ are as in Eq.~\eqref{eq:rho}. To simplify the optimization problem and lead to faster algorithms, note that we do not impose positivity of the distributions $\rho \in \mathcal{P}$, though this could be enforced, for instance, by imposing the linear constraints $\rho(R_i) \geq 0$ for a suitable choice of $R_i \in \operatorname{SO}(3)$. Moreover, the constraint $\int_{\operatorname{SO}(3)} \rho(R) dR = 1$ is equivalent to imposing $B_{0,0} = 1$ \cite{nonuniform}, which can be achieved by removing $B_{0,0}$ from the set of optimization variables and fixing its value to $1$. The values of the bandlimit parameters $L, P$ and the hyperparameter $\lambda$ used in our numerical experiments are given in~\Cref{app:hyperparameter}.

Just as in the previous section, we replace the continuous norms in Eq.~\eqref{eq:continuous_iKam} by discrete norms to define the metric between empirical moments and the moments from a 3-D structure as
\begin{align}\label{eq:d-Kam}
&d_{\operatorname{iKam}} \left( (\tilde{\mathbf{m}}_1, \tilde{\mathbf{m}}_2) , \Phi \right)  :=  \left(\min_{\rho \in \mathcal{P}}  \| \tilde{\mathbf{m}}_2 - \mathbf{m}_2 (\widehat{\Phi}, \rho) \|_{w_2}^2  +  \lambda  \| \tilde{\mathbf{m}}_1 - \mathbf{m}_1(\widehat{\Phi}, \rho) \|_{w_1}^2\right)^{1/2}.
\end{align}
 The cost function in Eq.~\eqref{eq:continuous_iKam} is rotationally invariant, in that it does not depend on the orientation of $\Phi$, since Eq.~\eqref{eq:rotate_moments} implies that 
 \begin{equation}
 \begin{split}
&\min_{\rho \in \mathcal{P}} \,\,\, \| \tilde{\mathbf{m}}_2 - \mathbf{m}_2 (\widehat{R\cdot\Phi}, \rho) \|_{L^2(\mathbb{R}^2 \times \mathbb{R}^2)}^2 \, + \,\, \lambda \, \| \tilde{\mathbf{m}}_1 - \mathbf{m}_1(\widehat{R\cdot\Phi}, \rho) \|_{L^2(\mathbb{R}^2)}^2 \\
&= \min_{\rho \in \mathcal{P}} \,\,\, \| \tilde{\mathbf{m}}_2 - \mathbf{m}_2 (\widehat{\Phi}, R^T\cdot\rho) \|_{L^2(\mathbb{R}^2 \times \mathbb{R}^2)}^2 \, + \,\, \lambda \, \| \tilde{\mathbf{m}}_1 - \mathbf{m}_1(\widehat{\Phi}, R^T\cdot\rho) \|_{L^2(\mathbb{R}^2)}^2 \\
&= \min_{\rho \in \mathcal{P}} \,\,\, \| \tilde{\mathbf{m}}_2 - \mathbf{m}_2 (\widehat{\Phi}, \rho) \|_{L^2(\mathbb{R}^2 \times \mathbb{R}^2)}^2 \, + \,\, \lambda \, \| \tilde{\mathbf{m}}_1 - \mathbf{m}_1(\widehat{\Phi}, \rho) \|_{L^2(\mathbb{R}^2)}^2 .
\end{split}
 \end{equation}
where the last equality follows because $R^T\cdot\rho$ lies in $\mathcal{P}$, since rotating a viewing angle distribution over $\operatorname{SO}(3)$ results in another another viewing angle distribution over $\operatorname{SO}(3)$.

At the cost of a solving the small linear system detailed in~\Cref{app:ls}, our method allows for the comparison between a stack of images and a resolved structure, without performing a 3-D reconstruction.
Furthermore, we precompute the least-squares matrices necessary for optimization, after which the distance function can be calculated in real-time. With sufficient storage and precomputation, the procedure is scalable to the entirety of the PDB.

In particular, $d_{\operatorname{iKam}}$ can be used in an efficient scheme to match a stack of synthetic images to the potentials of nearby PDB structures. By selecting a subset of the PDB database, one can efficiently compute $d_{\operatorname{iKam}} \left( (\tilde{\mathbf{m}}_1, \tilde{\mathbf{m}}_2) , \Phi \right)$ for each $\Phi$ in the subset and find the nearest neighbors.The method for processing image moments in practice is detailed in~\Cref{app:basis} and the computational complexity of the metric is derived in~\Cref{app:complexity}.

\section{Results} \label{sec:results}
\subsection{Existing measures of structure similarity}\label{sec:similarity_metrics}
There are several existing methods for reporting structure similarity between two known volumes. We list two approaches based on computing alignment and Zernike moments. We compare both $d_{\operatorname{iKam}}$ and $d_{\operatorname{vKam}}$ to these approaches in the experiments in the following subsections. Note that the following existing metrics are limited to measuring similarity between two structures and cannot compare images to structures, whereas $d_{\operatorname{iKam}}$ can.
\begin{enumerate}
    \item \textbf{Euclidean alignment}: A classical approach for comparing the similarity of two structures is to sample the volumes on a 3-D grid and calculate the Euclidean distance between the pair over rotations and translations. However, this method is expensive to compute since optimization over $\operatorname{SO}(3)$ is required to align the structures. Accelerated methods for computing these alignments by maximizing the correlation between two volume maps over rotations and translations have been implemented in various programs~\textit{e.g.} via gradient ascent in Chimera \cite{chimerax}. Further acceleration can be achieved by calculating volumetric correlations by expanding the volumes in a well-chosen basis and applying dimensionality reduction \cite{Rangan2023} or by maximizing the correlation between common lines in projection images generated from the volumes \cite{harpazshkolnisky2023}. Similar alignment methods, such as those described in \cite{Bartesaghi2008, XU2012152}, are also used in electron tomography for sub-volume similarity. In this paper, we use a Bayesian optimization algorithm to minimize a Euclidean loss function, as described in \cite{singer2023alignment}, to compute the alignment and minimum distance between two volumes. 
    \item \textbf{Zernike moments}: Another metric for structure similarity is to expand the molecule's structure in Zernike polynomials and compute a metric from the Zernike expansion coefficients, as described in \cite{guzenko2020real}, which is used by the PDB for structure similarity search.
\end{enumerate}

\subsection{Applying $d_{\operatorname{vKam}}$ to a PDB subset}
\label{sec:database}

To test the ability of $d_{\operatorname{vKam}}$ to discern the similarity between 3-D structures, we first generate a database using 1420 structures downloaded from the PDB~\cite{berman_protein_2000}. The subset chosen here was selected by filtering for human proteins with an experimental structure at resolution between 1-3\r{A} and a molecular weight between 150-250 kDa. We use this subset because it encompasses a diverse range of shapes and symmetries as well as many homologous structures. Additionally, the weight range reflects a smaller and more challenging protein size for a typical cryo-EM experiment~\cite{cianfrocco_what_2020}. In the future, a larger database containing the entire PDB can also be generated. 

\begin{figure}[!h]
	\includegraphics[width=0.8\textwidth]{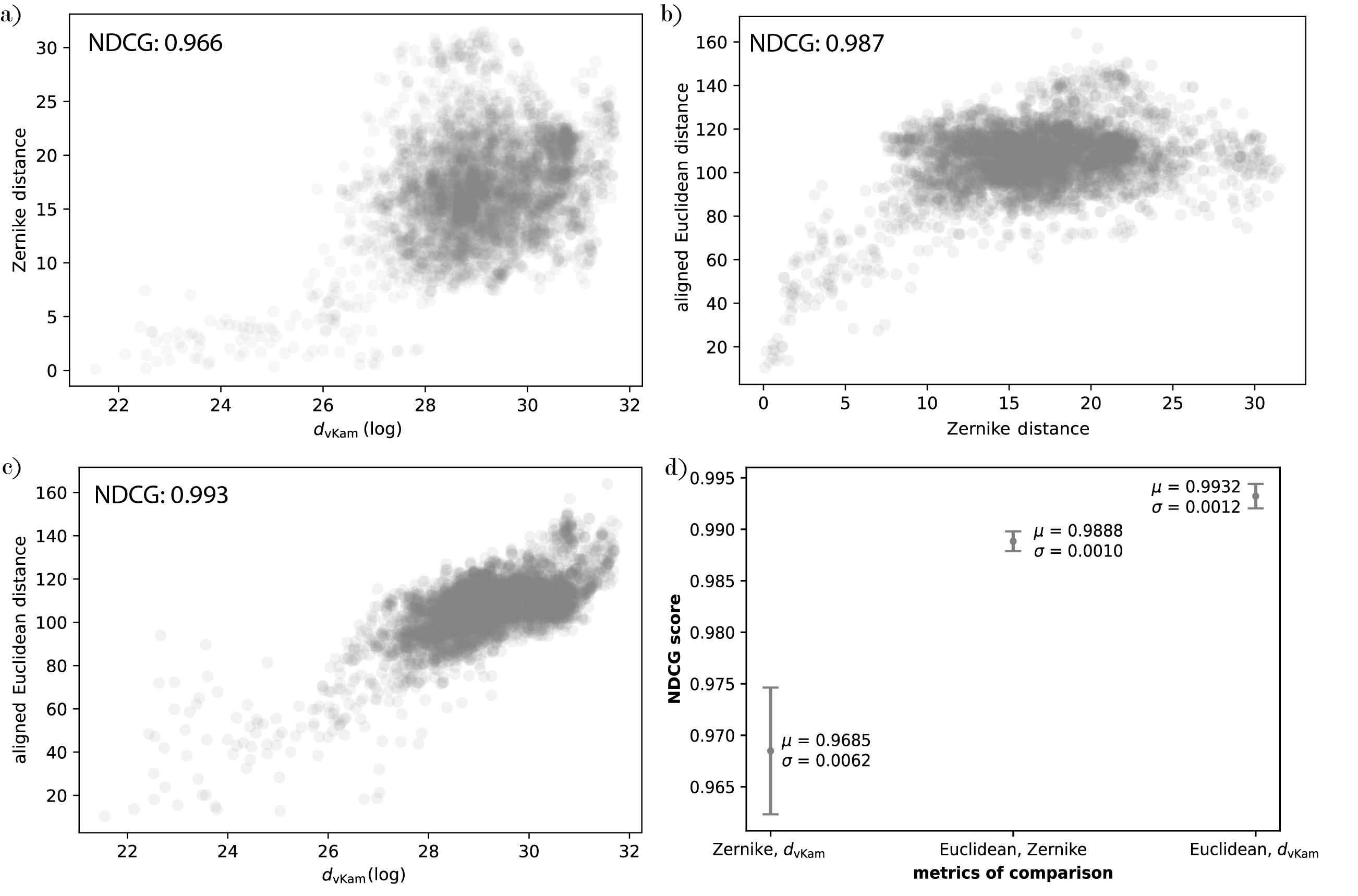}
	\centering
	\caption{Comparison between $d_{\operatorname{vKam}}$, the Zernike metric, and Euclidean alignment. a-c) a random size 100 subset of the database is selected. Then, pairwise similarity metrics are calculated and plotted, where each point represents a pair of structures. The NDCG score is calculated using the metric on the $y$-axis as the predicted metric, and the metric on the $x$-axis as the true metric. d) The procedure is repeated with 10 randomly selected size 100 subsets and the mean ($\mu$) and standard deviation ($\sigma$) of the NDCG scores are calculated. The errorbars and points visualize $\mu \pm \sigma$.}
	\label{fig:vkam-plots}
\end{figure}

Using our database, we first generate a discretized potential for each structure as described in~\Cref{app:uniform}. The first and second moments of each structure can then be computed using Eq.~\eqref{eq:moments}. We then compute $d_{\operatorname{vKam}}$ in Eq.~\eqref{eq:d-Kam1} pairwise for all structures in the database. 

To compare the performance of $d_{\operatorname{vKam}}$ against existing metrics, we calculate pairwise scores using $d_{\operatorname{vKam}}$, Euclidean alignment, and the Zernike metric. We then plot the returned scores against each other and calculate a ranking similarity using Normalized Discounted Cumulative Gain\cite{ndcg} (NDCG). We use this metric since it is a popular method to quantify the similarity between sets of rankings; its calculation is given in~\Cref{app:NDCG}.

In~\Cref{fig:vkam-plots}, we report the NDCG scores between pairs of metrics. All  NDCG scores are close to 1, indicating strong agreement among the three different metrics on which structures are most similar. However, the alignment metric and $\log(\text{$d_{\operatorname{vKam}}$})$ share the highest average NDCG score.
To verify the statistical signifance of this agreement, we report a t-test by selecting $10$ different subsets, showing that the NDCG score between $d_{\operatorname{vKam}}$ and the alignment metric is statistically significantly higher (with a $p$-value $p \approx 8 \times 10^{-9}$) than the NDCG score between the alignment metric and the Zernike metric. We thus conclude that $d_{\operatorname{vKam}}$ provides a fast and accurate alternative for the alignment metric.

Although it is the most interpretable metric, Euclidean alignment is computationally expensive to execute for all pairs of structures in a database. To achieve a manageable runtime for alignment, we calculate pairwise Euclidean alignment distances for subset of the database of size 100. Pairwise alignment on this subset took 8 hours on a 2.6 GHz Intel Skylake CPU running AVX-512 using 16 physical cores and 80 GB RAM. To do pairwise alignment via Bayesian optimization for the entire database of 1420 structures would require 46 days of computation, whereas using $d_{\operatorname{vKam}}$ (including precomputation) to calculate pairwise distances between all 1420 structures in the database requires 3 minutes on the same hardware. Despite containing an alignment component, the Zernike metric is also fast, taking 3 minutes to compute pairwise distances for the entire database.

\begin{figure}[!h]
	\includegraphics[width=0.6\textwidth]{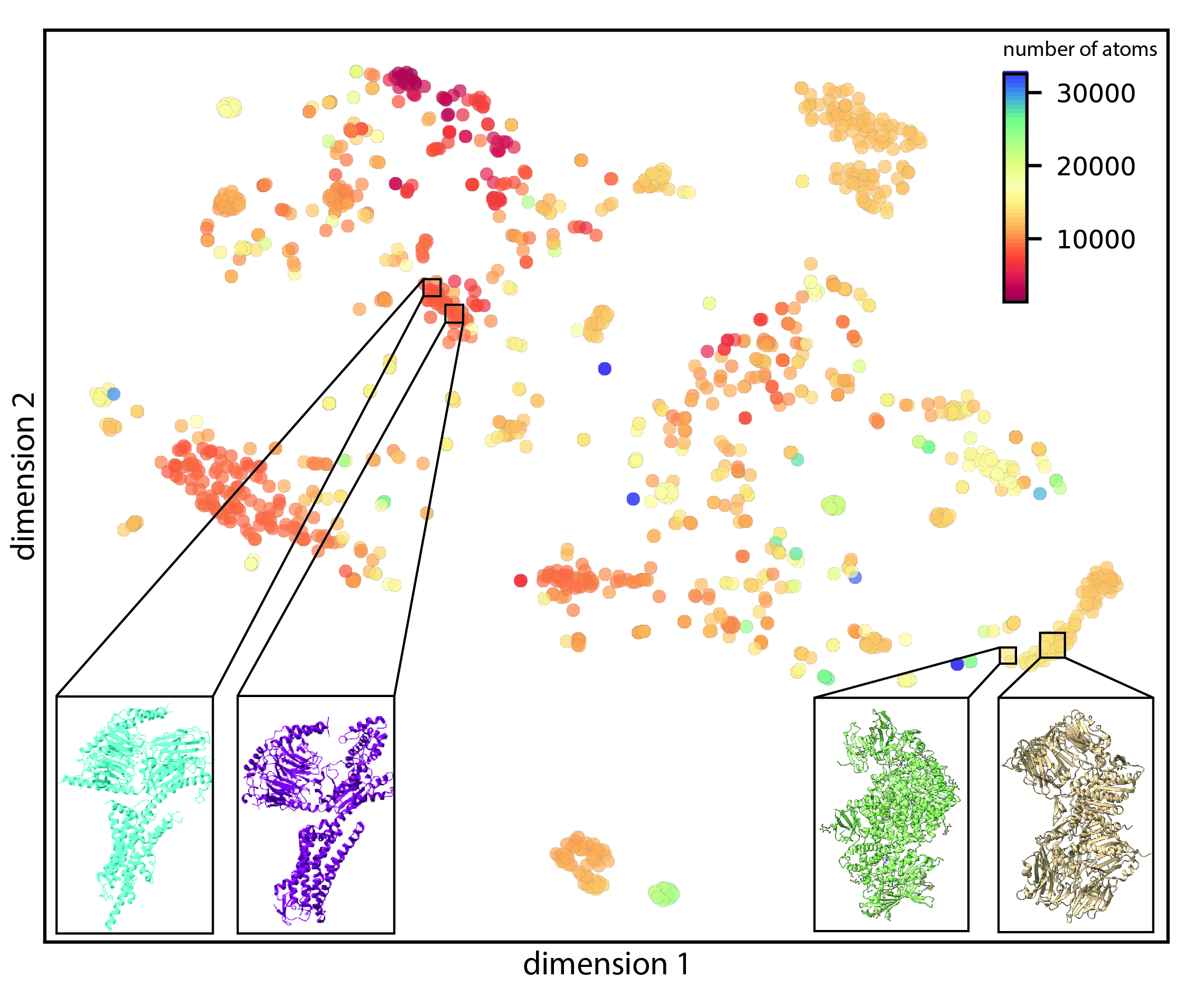}
	\centering
	\caption{2-D embedding of protein structures based on their similarity using $d_{\operatorname{vKam}}$. The analytical moments of 1420 proteins were computed and compared using Eq.~\eqref{eq:d-Kam1}, and t-SNE was applied for visualization. Each node represents a single structure and is colored by the number of atoms. Distinct clusters containing homologous or similarly shaped structures suggest that $d_{\operatorname{vKam}}$ provides interpretable results.}
	\label{fig:atlas}
\end{figure}

After observing high agreement between $d_{\operatorname{vKam}}$ and the other metrics, we compute a 2-D embedding of the similarity between structures in our database using t-SNE \cite{tsne} (see~\Cref{fig:atlas}). Analogous t-SNE plots for the alignment metric and Zernike metric are reported in~\Cref{app:tsne}. We find that $d_{\operatorname{vKam}}$ provides interpretable results in identifying similar molecules from their moments without the need for alignment. In particular, we observe that both homologous (\textit{i.e.}, structures with similar sequences) and similar-shaped structures are shown to be clustered together.

\subsection{Database search using $d_{\operatorname{iKam}}$ with synthetic cryo-EM data}
\label{sec:image-metric}
We next demonstrate the ability of $d_{\operatorname{iKam}}$ to accurately find a  match for the moments computed from projection images to a database of analytical moments computed from the atomic coordinates of known structures. To test our metric, we use the same dataset as the previous section, selecting the protein structure of a Mas-related G-protein-coupled receptor (available as entry PDB-7VV3 \cite{7VV3}) from our database described in ~\Cref{sec:database}.
We use this entry because our database includes several similarly shaped yet non-identical structures, on which we examine our metric's performance.

\begin{figure}[!h]
	\includegraphics[width=1\textwidth]{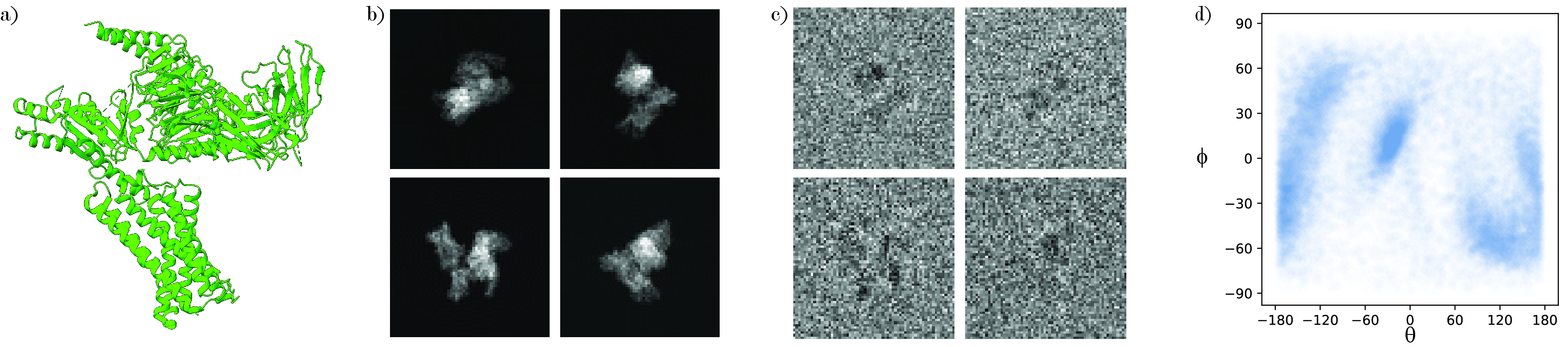}
	\centering
	\caption{Visualization of the generation of simulated images. (a) Protein structure of PDB-7VV3 (b) Clean projection images from PDB-7VV3 generated with a nonuniform viewing angle distribution (c) Projection images corrupted with a CTF and white noise with $SNR = 0.1$. (d) Distribution of nonuniform viewing angles}
	\label{fig:noisy}
\end{figure}

We generate a synthetic cryo-EM dataset as illustrated in~\Cref{fig:noisy}: we take $25000$ clean projection images from a nonuniform distribution over $\operatorname{SO}(3)$ at viewing angles given by a mixture of three von Mises-Fisher distributions\cite{spheredistributions}. To simulate cryo-EM data, the images are then corrupted with one of 100 unique radial CTFs, after which we add white noise with a signal-to-noise ratio (SNR) of $0.1$. We define the SNR by taking the signal as the average squared intensity over each pixel in all the clean images, and setting the noise variance to the appropriate ratio of the signal. These simulated images are generated using the ASPIRE software package \cite{aspire} and have parameters consistent with many experimental datasets. 

\begin{figure}[!h]
	\includegraphics[width=0.5\textwidth]{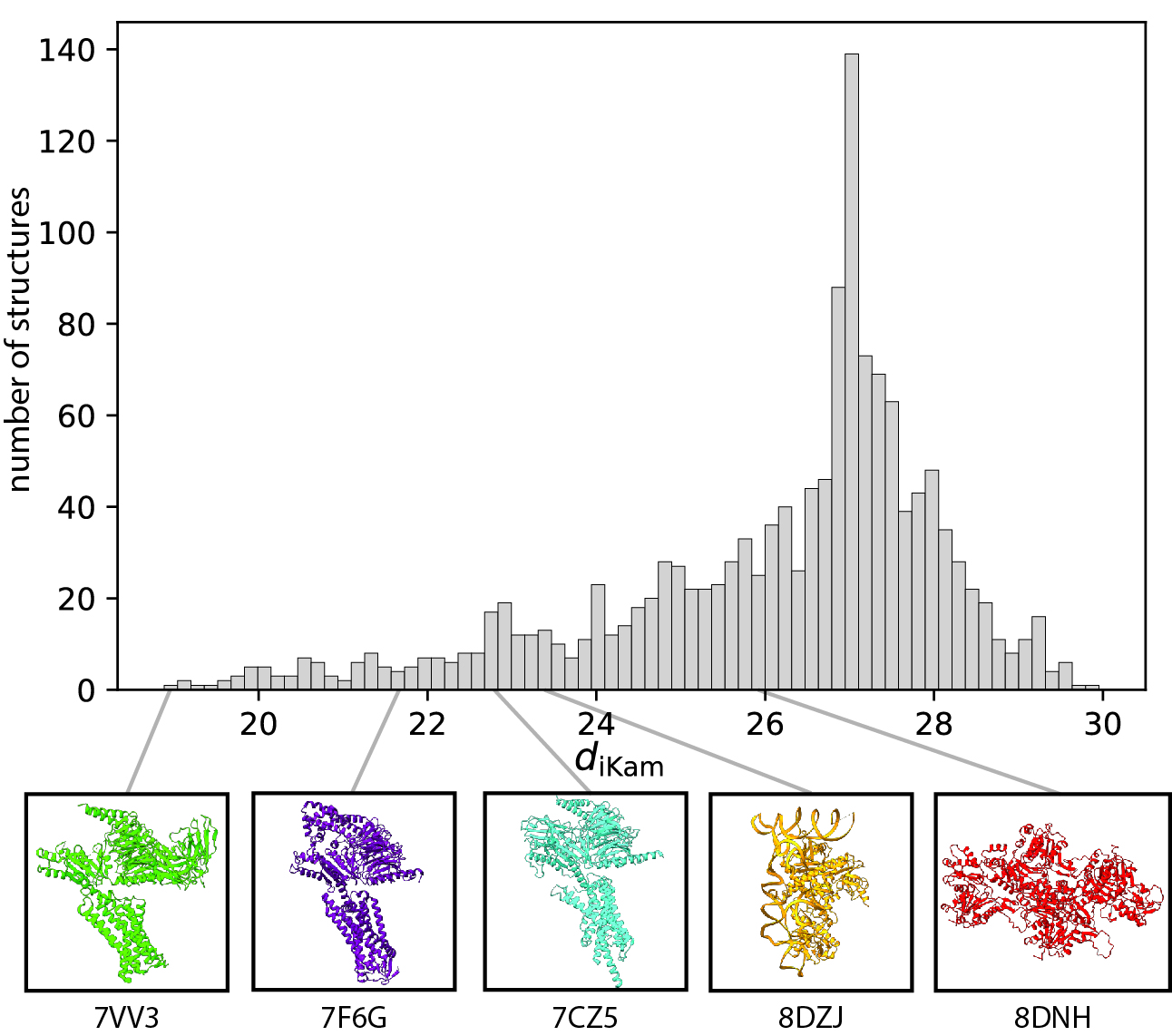}
	\centering
	\caption{Histogram ranking of dissimilarities computed using $d_{\operatorname{iKam}}$ on simulated noisy projection images generated from PDB-7VV3.}
	\label{fig:bar}
\end{figure}

We then compute the moments of the simulated images as will be shown in Eq.~\eqref{eq:first-moment} and \eqref{eq:second-moment} and compare to the database of moments using the image-to-volume metric described in Eq.~\eqref{eq:continuous_iKam}. We also report the effect of varying the number of images on the metric's performance in~\Cref{app:robustness}. Using our metric, we can rank the similarity of the image's moments to our database as shown in~\Cref{fig:bar}. We show that the most similar score (\textit{i.e.}, the smallest value in image Kam's metric) corresponds to the ground truth structure used to generate the images. Furthermore, based on our results, the next top 116 structures correspond to structures with similar volumes and sequences. 
These results demonstrate that we are able to compare directly between noisy, CTF-corrupted images and known structures. This approach could be especially valuable if there is no known model for initialization in 3-D reconstruction or if the molecule generating the images is unknown~\cite{verbeke_classification_2018}.

\begin{figure}[!h]
	\includegraphics[width=1\textwidth]{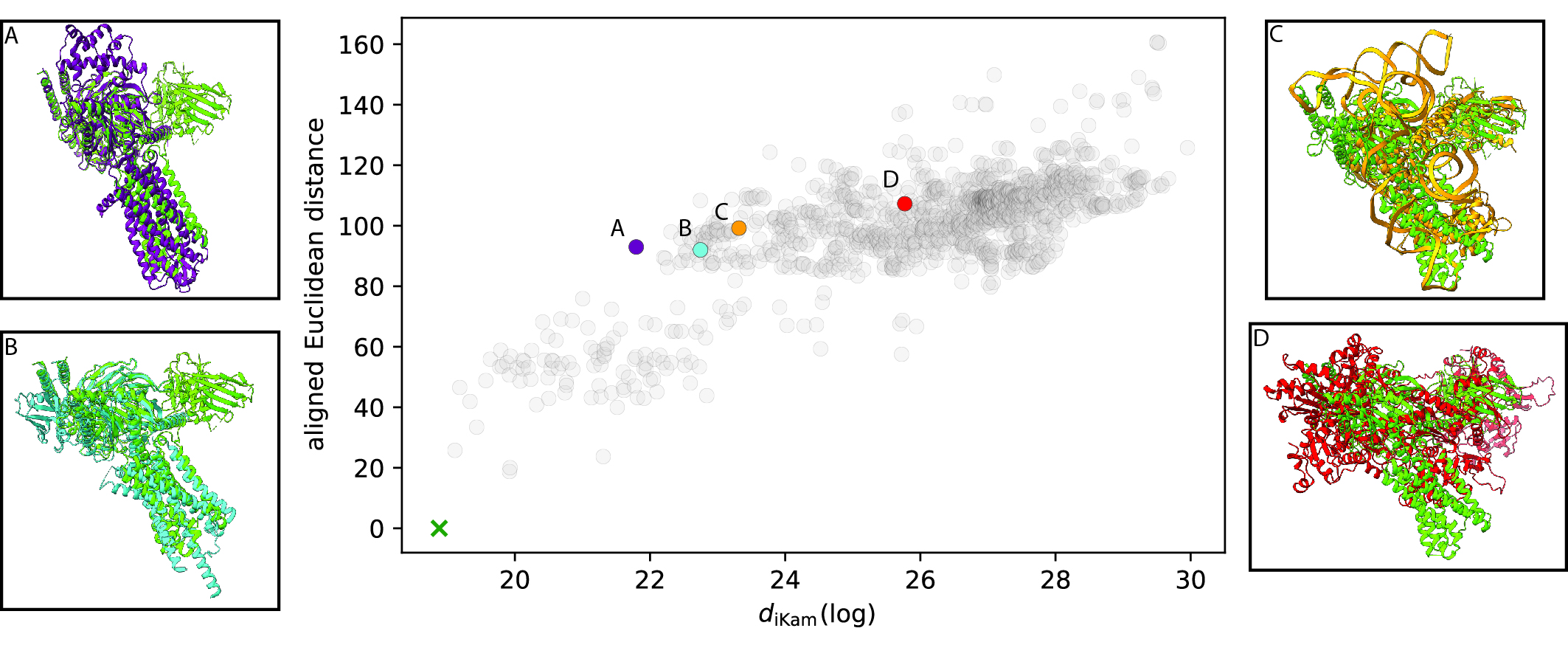}
	\centering
	\caption{Comparison between the rankings given by $d_{\operatorname{iKam}}$ (computed from simulated images) and the minimum Euclidean distance after alignment (computed from volumes). The structures shown are superimposed with the ground truth after alignment in panels A-D. The points on the graph that correspond to these structures are colored and labeled. The ground truth corresponds to the green cross in the lower left. Note that the Euclidean alignment metric shows stagnation whereas Kam's metric does not.}
	\label{fig:align}
\end{figure}

We report alignment scores between molecules in our database to PDB entry 7VV3, compare these to our metric's scores, and plot the results in \Cref{fig:align}. Most notably, when the protein structure becomes less similar to the ground truth (7VV3), the alignment metric begins to lose discriminative power. \Cref{fig:align} shows structures with varying degrees of dissimilarity as having the same score ($\sim$100). In contrast, our metric retains discriminative power, ranking structures with similar sequences/functions before structures with similar shapes.

Alignment via Bayesian optimization between one structure and the 1420 structures in the database took 95 minutes using the hardware described in~\Cref{sec:database}. Aside from the computational cost, the interpretation of the optimal rotation returned by alignment becomes unclear when comparing two structures that are not volumetrically similar. On the other hand, our metric does not return an alignment between two structures, which could render it less useful when an explicit alignment must be computed. Without this alignment, it may become harder to visually compare their volumes.

\begin{figure}[!h]
	\includegraphics[width=0.5\textwidth]{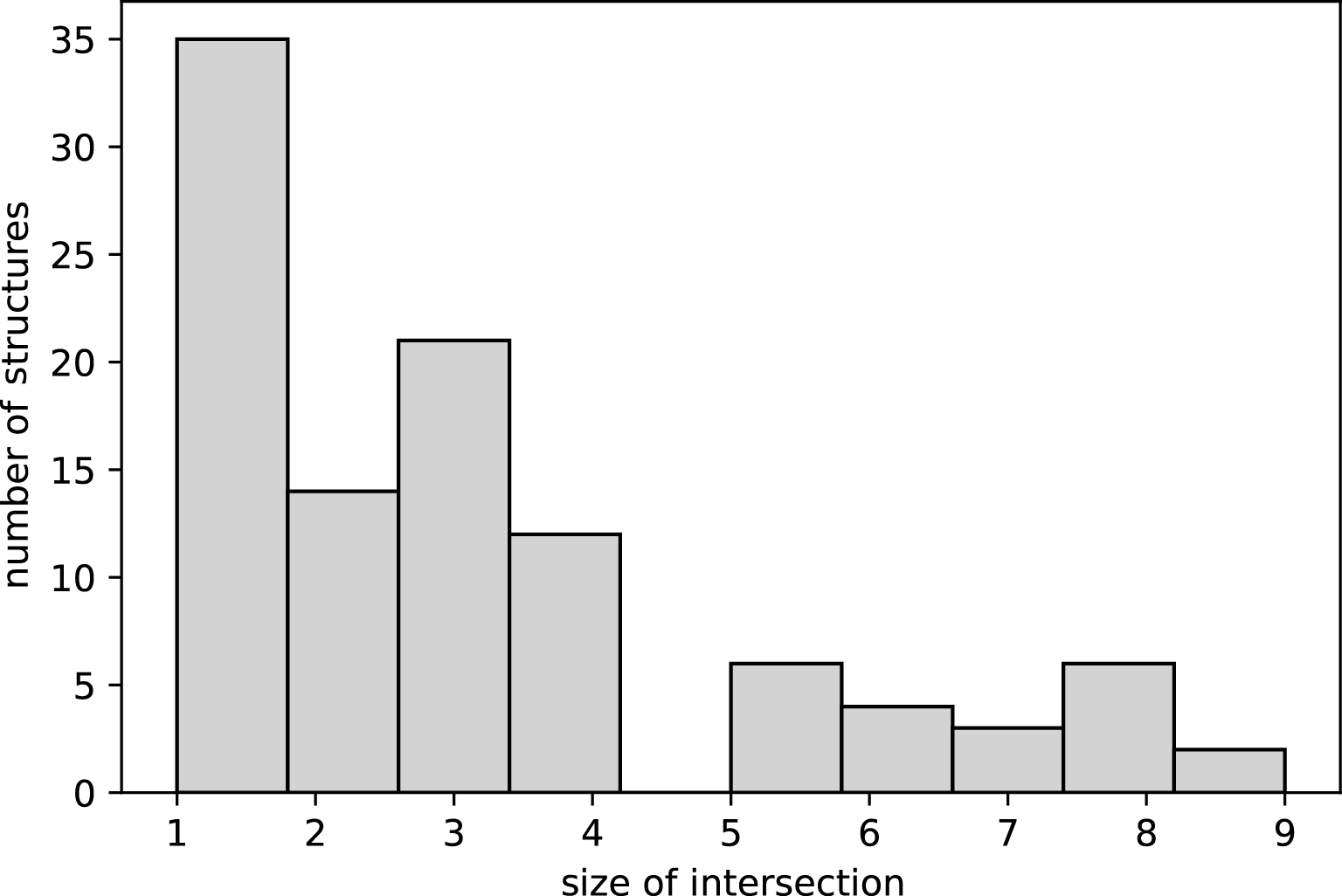}
	\centering
	\caption{Comparison between $d_{\operatorname{iKam}}$ computed from simulated images, and the Zernike metric, computed from volumes. Here, we repeat simulated experiments 100 times. Then, the size of the intersection of the top ten structures returned by $d_{\operatorname{iKam}}$ and the Zernike metric is plotted as a histogram.}
	\label{fig:overlap}
\end{figure}

It is computationally costly to generate and perform moment estimation on synthetic images for every molecule in the database. As such, to compare the performance of our metric against the Zernike metric, we select from our database a random subset of 100 structures. For each structure, we repeat the process we perform on PDB-7VV3: first, we generate a nonuniform distribution over $\mathbb{S}^2$ as a mixture of 3 von Mises-Fisher distributions with random means, weights, and covariance matrices. We then generate 25000 images, corrupt with SNR = 0.1 and radial CTFs, compute the moments, and search across the database.

For every structure, we recover the ground truth as one of the first six lowest-scoring molecules. Moreover, 88 of the 100 tests recovered the ground truth as the lowest-scoring molecule. To evaluate how well the metrics agree on structure similarity, we compute the size of the intersection between the top ten structures returned by our metric and those returned by the Zernike metric. As shown in~\Cref{fig:overlap}, we find that the metrics agree on two to three structures, and a large number of structures agree only on the ground truth structure. When they occur, disagreements between the metrics are likely due to the presence of near-identical molecules in the database.

\subsection{Towards matching experimental datasets by $d_{\operatorname{iKam}}$}\label{sec:experimental}

While our simulated result shows success in matching a synthetic cryo-EM dataset to PDB structures, many experimental cryo-EM datasets are corrupted by a large number of unmodeled effects that we have not considered. Among the real-data effects are: scattering potential's corruption by a solvent effect \cite{shang2012hydration}, the B-factor ~\cite{rosenthal2003optimal}, a global scaling ambiguity, imperfect centering, junk particles, non-radial CTF, and imperfect noise model.  Our simulation falls short on these counts.

In a first step towards applying $d_{\operatorname{iKam}}$ to real experimental datasets, we compare the moments of a stack of images deposited in the Electron Microscopy Public Image ARchive (EMPIAR) \cite{empiar} to the moments of its preprocessed 3-D reconstructions given by the program CryoSPARC~\cite{punjani2017cryosparc}. We select the dataset EMPIAR-10076~\cite{davis2016ribosome}, a heterogeneous dataset containing five major structures. The dataset is well characterized, and each image in the dataset has been classified to one of the five major states~\cite{davis2016ribosome} or ``junk" particles, which we discard. We use the classification to generate five separate datasets, allowing us to compute five different moments, one for each of the major states. 
This test case allows us to examine our metric matching on a real dataset, while bypassing some of the issues associated with comparing datasets and volumes obtained in different experimental conditions.

We downsample the image stack to $64 \times 64$, center using the deposited shift, and mask the images with a circular binary mask of radius $0.8$ times half the side length of the image.
We then estimate the moments for each structure and compare them to moments computed analytically from preprocessed volume reconstructions of the five major structures, as well as two other distinctly-shaped ribosomes from the Electron Microscopy Data Bank\cite{emdb} (EMDB), EMD-8457 and EMD-2660, used as a baseline. Scaling issues between the moment computed from the images and the moment computed from the volume are resolved by examining the diagonal entries of the second moments. Specifically, we find a multiplicative scaling factor that best matches the diagonal of the image-computed second moment and those of the volume-computed second moment under a uniform distribution with respect to the $l^2$-norm.

\begin{figure}[!h]
	\includegraphics[width= 1 \textwidth]{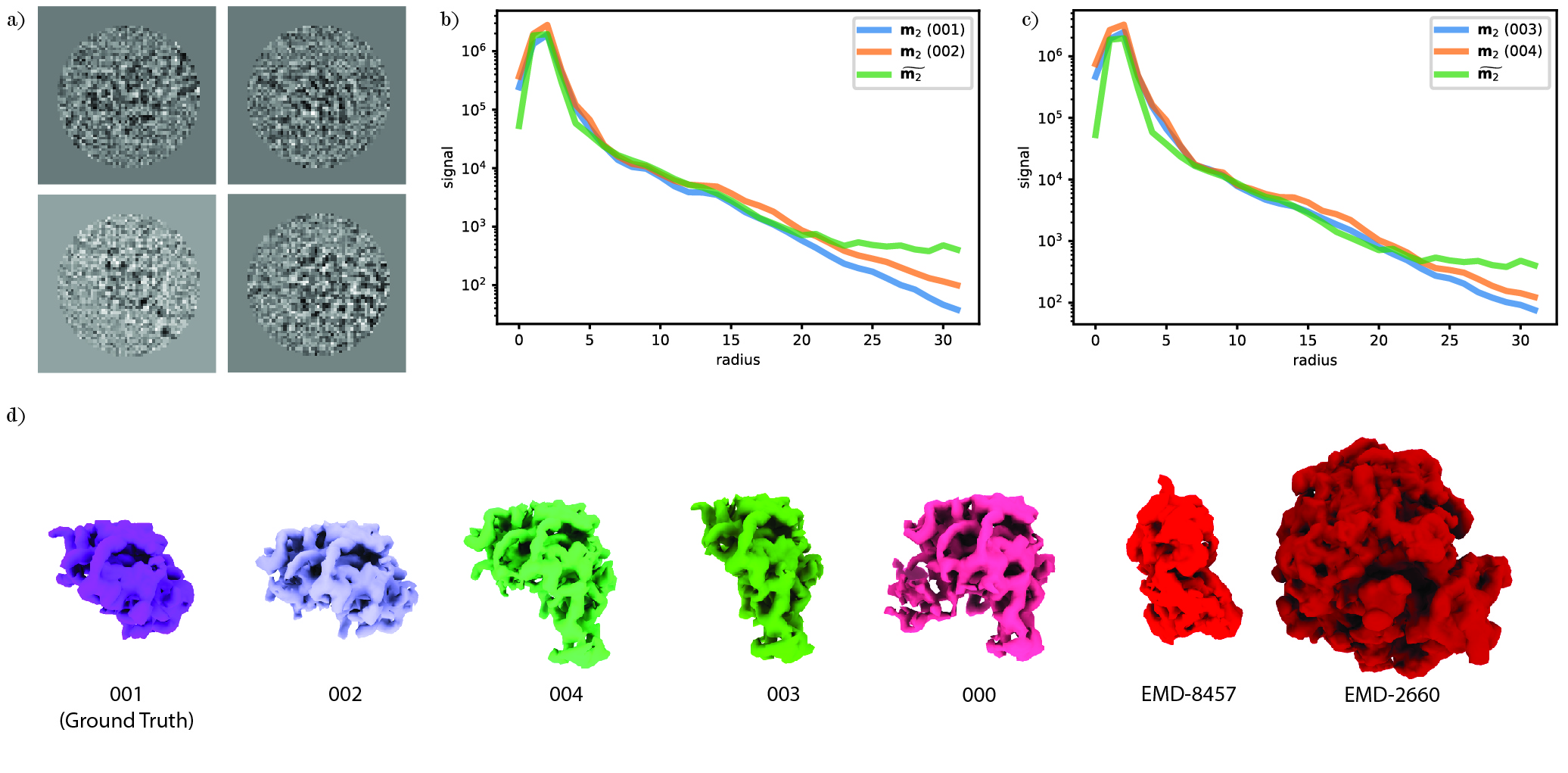}
	\centering
	\caption{$d_{\operatorname{iKam}}$ visualization and ranking results for experimental data corresponding to structure 001 (a) Experimental images from EMPIAR-10076 corresponding to structure 001 downsampled to $64 \times 64$ pixels, centered, and with binary mask applied. (b) Comparison between diagonal entries of the second moment computed from the reconstructed volumes 001, 002 and the moment estimated from experimental images corresponding to structure 001. (c) Comparison between diagonal entries of the second moment computed from the reconstructed volumes 003, 004 and the moment estimated from experimental images from experimental images corresponding to structure 001. (d) The five reconstructions (000-004) and two baseline structures (EMD-8457, EMD-2600) ranked using $d_{\operatorname{iKam}}$, ordered from left to right.
 }
	\label{fig:experimental}
\end{figure}
\begin{figure}[!h]
	\includegraphics[width= 1 \textwidth]{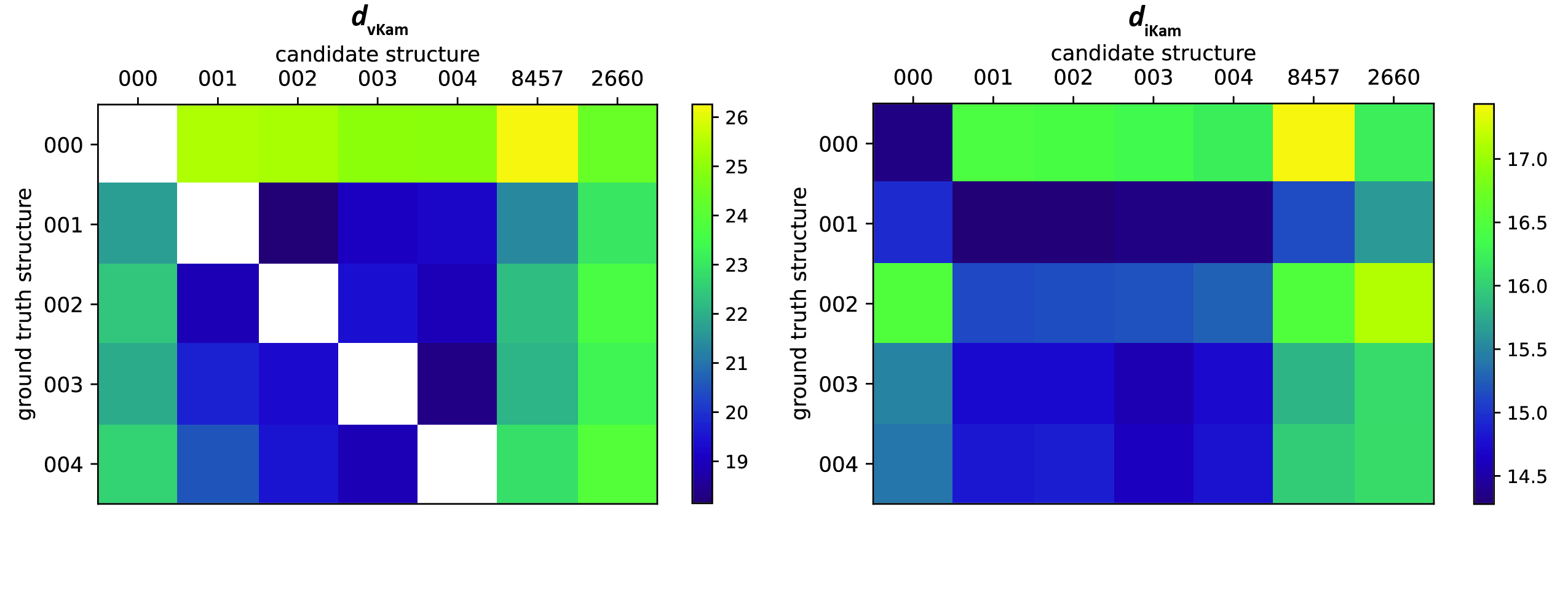}
	\centering
	\caption{Visualization of  $\log(d_{\operatorname{vKam}})$ and $\log(d_{\operatorname{iKam}})$ values on the seven candidate structures. Here, EMD-8457 and EMD-2660 are listed as 8457 and 2660 for brevity. Note that there are five ground truth structures but seven candidate structures since EMD-2660 and EMD-8457 are baseline structures for which there are no images in the experimental dataset. }
	\label{fig:metric-heatmap}
\end{figure}

As shown in~\Cref{fig:experimental}, it is observed that Kam's metric recovers the ground truth structure at the lowest distance for the experimental images corresponding to structure 001. We note that the scores for molecules 001 and 002, as well as molecules 003 and 004 are almost identical in value. Also, we find that the analytical moments are closer to each other than to the experimentally determined moments. Finally, the metric reports the baseline structures, which are very different in shape and size, at the largest distances.

In~\Cref{fig:metric-heatmap} we plot the distances between the five reconstructions (or in the case of $d_{\operatorname{iKam}}$, their experimental images) and the seven candidate structures given by both of our metrics. The exact values for $d_{\operatorname{iKam}}$ are given in~\Cref{app:experimental}. There is also scaling ambiguity in $d_{\operatorname{vKam}}$ since our reconstructions are preprocessed, hence we use the same approach as above: we scale each candidate structure's moment by a multiplicative scaling factor that best matches the candidate structure's diagonal entries of the second moment with those of the ground truth structure. 
Analyzing the trends in each row, we observe that the metrics seem to agree on the general ranking of the molecules. While the structures 001, 002 and 003, 004 are very similar, $d_{\operatorname{vKam}}$ shows that the metric distinguishes between them given accurate moment estimation, whereas $d_{\operatorname{iKam}}$ loses some discriminative power. However, when it comes to distinct molecules such as EMD-8457 and EMD-2660, both metrics agree on their rankings.

\section{Limitations and Future Work} \label{sec:limitations}
$d_{\operatorname{iKam}}$ currently falls short of being directly applicable to experimental datasets. 
As stated in \Cref{sec:experimental}, there are several unmodeled effects not considered in this work that could lead to unexpected results for real data. The net effect of ignoring these experimental considerations is to bias our moment estimator, which may explain the inability of $d_{\operatorname{iKam}}$ to detect the smaller differences between structures 001 and 002, as well as 003 and 004. Developing an estimator that is robust to outliers (such as junk particles) could help alleviate this. 

While we address a few of these parameters, we do so with prior knowledge. For example, the shifts used to center images are a byproduct of the reconstruction process. In future work, we aim to develop methods to correct for these effects directly from the raw images. Likewise, here we have controlled for experiment-specific artifacts by using images and structures resolved from the same experiment, whereas in the future we wish to compare across all structures. 
Furthermore, in the future we seek to compare moments computed from real data directly to the PDB, by appropriately correcting for the discrepancies between PDB and reconstructed structures. 

Even with our current mitigations, issues such as the B-factor and inaccuracies in the noise model remain completely unmodeled. Further studies will be required to investigate which of these omissions is important and which can safely be made. Then, our method could be modified to account for the important effects.

\section{Discussion}
We introduced structural similarity metrics for proteins based on moments, inspired by the moment computation in Kam's method.
$d_{\operatorname{vKam}}$ compares known 3-D structures according to the difference between the moments of their  potentials. 
We showed that the metric accurately captures similarity according to the rotationally-aligned Euclidean metric, an interpretable but expensive-to-compute molecular similarity metric. 
Therefore, $d_{\operatorname{vKam}}$ allows for the efficient comparison of large number of known structures.
A potential application is to improve the similarity search presently in the PDB, which uses the Zernike metric - a fast but less principled metric which involves learning weights and which our results suggest performs worse than ours.

A second metric, termed $d_{\operatorname{iKam}}$, allows for the computation of a similarity score between an unknown structure present in a large cryo-EM dataset and a solved structure. 
The computation of this metric does not require a 3-D reconstruction process for the image stack, and therefore is very efficient.
We showed on simulated projection images that our method could discrimate between even very similar proteins with reasonably sized datasets. 
If it were to work on experimental datasets, $d_{\operatorname{iKam}}$ could become a versatile tool for 3-D reconstruction. Typical reconstruction algorithms used in practice are only locally optimal, and thus require good initialization, which $d_{\operatorname{iKam}}$ could provide by returning the homologous structures present in the PDB. By extending the database to the entirety of the PDB and including structure predictions, both solved and predicted structures could be quickly compared against. 

Beyond its application to experiments, $d_{\operatorname{iKam}}$ demonstrates that Kam's method is a feasible strategy for high-resolution reconstruction. Recent works have improved the viability of Kam's method by using sparsity~\cite{bendory2023autocorrelation} or neural network~\cite{neuralnets} priors; likewise, the search over the PDB using Kam's metric can be interpreted as simply running Kam's method under a very strong prior, where only a finite number of structures appear with non-zero probability. Our results suggest that, if one could formulate a tractable prior over the manifold of proteins, Kam's method could yield high-resolution reconstructions.

\section*{Acknowledgments}
  J.K. and M.A.G. thank Bronson Zhou for helpful conversations.

\section*{Competing Interests} The authors declare no competing interests exist.

\section*{Authorship Contributions} 
All authors conceived of the project and designed the algorithms.  A.Z. wrote the software and performed the experiments.  All authors wrote the manuscript and approved its submission.

\section*{Funding Statement} 
Part of this research was performed while authors J.K., E.J.V., N.F.M., M.A.G., and A.S.
were visiting the long program on Computational Microscopy at the Institute for Pure and Applied Mathematics, which is supported by NSF DMS 1925919.
A.S., A.Z., O.M., E.J.V., M.A.G. are supported in part by AFOSR FA9550-20-1-0266, the Simons Foundation Math+X
Investigator Award,  NSF DMS 2009753, and NIH/NIGMS  R01GM136780-01. J.K. is supported in part by NSF DMS 2309782, NSF CISE-IIS 2312746, and start-up grants from the College of Natural
Science and Oden Institute at the University of Texas at Austin. 
N.F.M. is supported in part by a start-up grant from Oregon State University.  

\section*{Data Availability Statement} Replication code can be found at \url{https://github.com/aszhang107/moment-based-metrics/}.

\bibliographystyle{unsrtnat}
\bibliography{references.bib}

\appendix
\section{Methodology} \label{sec:methodology}In this section, we describe the computational details of the method.

\subsection{Moment derivation}\label{app:moments}
Prior work \cite{nonuniform} has shown that the analytical first and second moments of cryo-EM images generated by $\widehat{\Phi}$ and $\rho$ equal
\begin{align} \label{eq:first-moment}
& \mathbf{m}_1 (\widehat{\Phi}, \rho) = \sum_{\ell, m} B_{\ell, -m} A_{\ell, m}(r) N^{0}_{\ell} \frac{1}{2 \ell + 1} (-1)^m,
\end{align}
where the sum ranges over $(\ell, m)$ such that $0 \leq \ell \leq \min(L,P)$, $\ell$ is even, and $-\ell \leq m \leq \ell$, and 
\begin{multline} \label{eq:second-moment}
\mathbf{m}_2  (\widehat{\Phi}, \rho) = \sum_{n} e^{i n(\phi' - \phi)} \\ \sum_{\substack{\ell, m,  \ell', m', \ell''}} \!\!\! A_{\ell, m}(r) A_{\ell', m'}(r') N^{n}_{\ell} N^{-n}_{\ell'} B_{\ell'', -m-m'} C_{\ell''}(\ell, \ell', m, m', n, -n) \frac{(-1)^{m+m'}}{2\ell''+1},
\end{multline}
where \begin{equation}
N^n_{\ell} := \begin{cases} \sqrt{ \frac{2 \ell + 1}{4 \pi} \frac{(\ell - n)!}{(\ell + n)!} } \frac{(-1)^n}{2^{\ell} \ell!} (\ell + n)! \binom{\ell}{(\ell + n)/2} (-1)^{(\ell-n)/2} & \text{ if } n \equiv \ell \, (\text{mod }2) \\
0 & \text{if } n \not \equiv \ell \, (\text{mod } 2)
\end{cases}
\end{equation}
is an explicitly calculated constant,
\begin{equation}
C_{\ell''}(\ell, \ell', m, m', n, n') := C(\ell, m; \ell', m' \vert \ell'', m+m') C(\ell, n; \ell', n' \vert \ell'', n+n') 
\end{equation} is a product of Clebsch-Gordan coefficients \cite{angularmomentum}, and the sum ranges over those indices $n,\ell, m, \ell', m', \ell''$ that satisfy
\begin{equation}\label{eq:gnarly_summation_ranges}
\begin{split}
&0 \leq n \leq L, \quad -n \leq \ell \leq n, \quad -n \leq  \ell' \leq n, \quad -\ell \leq m \leq \ell, \quad -\ell' \leq m' \leq \ell',\\
&\ell \equiv \ell' \equiv n \!\!\! \mod 2, \quad \max(|\ell - \ell'|,|m + m'|) \leq \ell'' \leq \min(\ell + \ell', P).
\end{split}
\end{equation}
See Sections 2.3.1 and 2.3.2 in \cite{nonuniform} respectively for the derivations of \eqref{eq:first-moment} and \eqref{eq:second-moment}.
In the case of the uniform density on $\operatorname{SO}(3)$, we note that $N_0^0 = \frac{1}{\sqrt{4\pi}}$ so the Eq.~\eqref{eq:first-moment} and \eqref{eq:second-moment} simplify to the following:
\begin{align} \label{eq:first-moment-unif}
\mathbf{m}_1(r) &= \sqrt{\frac{1}{4\pi}}A_{0, 0}(r), \\
\begin{split}
\mathbf{m}_2(\Delta \phi, r, r') &= \frac{1}{4\pi}\sum_{\ell, m} A_{\ell, m}(r) A_{\ell, -m}(r') P_\ell(\cos(\Delta \phi + \pi)) \\
&= \frac{1}{4\pi} \sum_{\ell, m} A_{\ell, m}(r) \overline{A_{\ell, m}(r')} P_\ell(\cos(\Delta \phi)), \label{eq:second-moment-unif}
\end{split}
\end{align}
where $P_\ell$ is the Legendre polynomial of degree $\ell$ and $\overline{z}$ denotes the complex conjugate of a complex number $z \in \mathbb{C}$. 
The simplification of \eqref{eq:first-moment} to \eqref{eq:first-moment-unif} is immediate, whereas the simplification of \eqref{eq:second-moment} to \eqref{eq:second-moment-unif} uses the sum rule for spherical harmonics, see Eq.~10 in \cite{Kam}.

\subsection{Uniform case}
\label{app:uniform}
This section details the method to compute $d_{\operatorname{vKam}}$.
Our algorithm takes as input a PDB identifier (a list of atomic coordinates), on which we center the atomic positions by subtracting the molecule's center of mass. Then we use the three-dimensional non-uniform fast Fourier transform (NUFFT) \cite{barnett2019parallel,barnett2021aliasing} to compute the discrete Fourier transform evaluated on a grid in spherical coordinates, i.e., to compute 
\begin{equation}\label{eq:discrete_fourier_spherical}
a_{kjl} = \sum_{i=1}^{q} \widehat{f}_i(r_k \alpha_{jl}) e^{\imath x_i \cdot r_k \alpha_{jl}} , \text{ where } \alpha_{jl} = (\sin\theta_j\cos\varphi_l,\sin\theta_j\sin\varphi_l,\cos\theta_j),
\end{equation}
where $x_i$ denotes the coordinates of the $i^\text{th}$ atom from the PDB identifier and $q$ is the total number of atoms. The function $\widehat{f}_i$ is the Fourier transform of the scattering potential of the $i^\text{th}$ atom as reported in \cite{electron_scattering, wilson_singer}. In real space, this corresponds to convolving a Gaussian mixture with a delta function, in other words, adding a Gaussian blob around the atom coordinate. 
Here, $r_k = \frac{k\delta}{N}$, $\theta_j = \frac{\pi j}{N}$ and $\varphi_l = \frac{2\pi l}{N}$ for $k = 0, \ldots, N/2$ and $j,l=0, \ldots, N-1$ and $\delta$ is the side length of the volume grid in angstroms. 

Lastly, we apply the spherical harmonic transform to $a_{kjl}$ defined on the spherical coordinate grid $(r_k, \theta_j, \varphi_l)$ in Eq.~\eqref{eq:discrete_fourier_spherical} using SHTools \cite{driscoll1994computing,wieczorek2018shtools}. This gives us coefficients 
$$
A_{\ell, m}(r_k) = \sum_{\substack{0 \leq \ell \leq L \\ 0 \leq k \leq n-1}}a_{k,j,l}\overline{Y_\ell^m(\alpha_{jl})}.
$$
 Let $\rho_u$ denote the uniform density on the sphere. In the discrete case, we sample each image as a 2-D polar grid at $N / 2$ radial points $r$ and $N$ angular points $\phi$, where $N$ is the number of pixels of one side of the projection images. In Eq.~\eqref{eq:first-moment}, the first moment $\mathbf{m}_1$ is indexed by $r$, and is thus an $N/2$-length vector. 
 Note that in Eq.~\eqref{eq: discrete_vKam}, $\Delta\phi_j = 2\pi j / N - 2\pi$ for $j = 1, \ldots , 2N$,
 but since $e^{in\Delta\phi}$ in Eq.~\eqref{eq:second-moment} is $2\pi/n$ periodic, we have that $e^{in(2\pi - \Delta\phi)} = e^{-in\Delta\phi}$ and hence $\mathbf{m}_2(r, r', \Delta\phi_j) = \mathbf{m}_2(r, r', \Delta\phi_{j+N})$. Thus, $\Delta\phi_{j}$ for $j = 1, \ldots, N$ is redundant and we consider only $\Delta\phi_{j}$ for $j = N + 1, \ldots, 2N$, which enumerates $[0, 2\pi]$. Thus, $\mathbf{m}_2$ is a three dimensional tensor of size $N \times N/2 \times N/2$, since there are $N$ values for $\Delta \phi $ and $N/2$ values each for $r$ and $r'$, where $\Delta \phi$ are points uniformly spaced between 0 and $2\pi$ and $r$, $r'$ are $N$ uniformly spaced points between 0 and $\delta$. Equations
\eqref{eq:first-moment-unif} and 
\eqref{eq:second-moment-unif}  give
\begin{align} 
\mathbf{m}_1(k) &= A_{0, 0}(r_k), \\
\begin{split}
\mathbf{m}_2(j,k_1,k_2) &= \frac{1}{4\pi}\sum_{\ell, m} A_{\ell, m}(r_{k_1}) A_{\ell, -m}(r_{k_2}) P_\ell(\cos(\Delta \phi_j + \pi)) \\
&= \frac{1}{4\pi} \sum_{\ell, m} A_{\ell, m}(r_{k_1}) \overline{A_{\ell, m}(r_{k_2})} P_\ell(\cos(\Delta \phi_j)).
\end{split}
\end{align}
We then compute the metric given in Eq.~\eqref{eq:d-Kam}. To better approximate the $L_2$-norm in the continuous case, we scale the difference of each entry $\mathbf{m}_2(j, k_1, k_2)$ by $\sqrt{r_{k_1} r_{k_2}}$ so that the squared norm is scaled by $r_{k_1} r_{k_2}$. More precisely, we define weighted $\ell^2$-norms $\|\cdot\|_{w_1}$ and $\|\cdot\|_{w_2}$ on $\mathbb{R}^{N/2}$ and $\mathbb{R}^{N \times N/2 \times N/2}$, as described in Eq.~\eqref{eq: discrete_vKam}. Let $\Phi$ and $\Phi'$ be two different molecules, and $\mathbf{m}_1, \mathbf{m}_2$ and $\mathbf{m}_1',\mathbf{m}_2'$ be the first and second moment tensor, respectively, from two different molecules. We define the distance between the moments as in Eq.~\eqref{eq:d-Kam1}.

\subsection{Least squares for the nonuniform case}\label{app:ls}
This section describes the process for generating and solving the least squares system for $B$, the matrix encoding the viewing angle distribution. We use the following convention for the vectorization operator $\text{vec}(\cdot)$: if $\mathcal{M} \in \mathbb{C}^{i \times j}$, $\text{vec}(\mathcal{M})$ returns a vector of dimension $ij$ obtained by stacking the columns of $\mathcal{M}$, i.e.,
\begin{equation}
    \text{vec}\left( \begin{bmatrix}  a_{1,1} & \cdots & a_{1,j} \\
    \vdots & \ddots & \vdots \\
    a_{i,1} & \cdots & a_{i,j}\end{bmatrix}\right) :=  \begin{bmatrix}  a_{1,1} & \cdots & a_{i,1} & a_{1,2}
    & \cdots & a_{i,j}\end{bmatrix}^T.
\end{equation}

The first moment is linear in $B$ as shown in Eq.~\eqref{eq:first-moment}, so fitting a viewing distribution to observed moments can be solved through a least-squares problem. We detail this procedure in \Cref{alg:cap}. \\
\begin{algorithm}
    \caption{Computation of least-squares matrix $V$ for $\mathbf{m}_1$}\label{alg:cap}

initialize $V[i = 1, \ldots, N/2][(p = 1,\ldots, P; m = -p, \ldots, p)] \gets 0$\\
\For{$i = 1,\ldots N/2$}{
    \For{$p = 1, \ldots, P$}{
        \For{$m = -p, \ldots, p$}{
        $V[i][(p;-m)] \gets A_{\ell, m}(r_i) N_\ell^0 \frac{(-1)^m}{2\ell + 1}$
        }
    }
}
\Return $V$

\end{algorithm}
\begin{algorithm}
\caption{Computation of least-squares matrix $\mathcal{U}^n_{\ell, \ell'}$ for $\mathcal{B}^n_{\ell, \ell'}$}\label{alg:BLSalg}
Initialize $\mathcal{U}^n_{\ell, \ell'}[i = 1, \ldots, (2\ell + 1)(2\ell' + 1) ][(\ell'' = 1,\ldots, P; m = -\ell'', \ldots, \ell'')] \gets 0$ \\
\For{$i = 1,\ldots (2\ell + 1)(2\ell' + 1)$}{
    \For{$m = -\ell, \ldots, \ell$}{
        \For{$m' = -\ell', \ldots, \ell$}{
            \For{$\ell'' = \max(|\ell - \ell'|,|m + m'|), \ldots, \min(\ell' + \ell, P)$}{
                $\mathcal{U}^n_{\ell, \ell'}[i][(\ell'', -m - m')] \gets \mathcal{U}^n_{\ell, \ell'}[i][(\ell'', -m - m')] + C_{\ell''}(\ell, \ell', m, m', n, -n) \frac{(-1)^{m+m'}}{2\ell''+1}$            }
        }
    }
}
\Return $\mathcal{U}^n_{\ell, \ell'}$

\end{algorithm}

For the second moment, we rewrite Eq.~\eqref{eq:second-moment} more compactly:
\begin{equation}\label{eq:m2_compact}
\mathbf{m}_2(\Delta \phi) = \sum_{n} e^{in\Delta \phi} \sum_{\ell, \ell'} \mathcal{A}_{\ell} \mathcal{B}^{n}_{\ell, \ell'} \mathcal{A}_{\ell'}^*,
\end{equation}
where
\begin{equation} 
\left(\mathcal{B}^{n}_{\ell, \ell'}\right)_{m, m'} = \sum_{\ell''} B_{\ell'', -m-m'} \mathcal{C}_{\ell''}(\ell, \ell', m, m', n, -n)\frac{(-1)^{m + m'}}{2\ell'' + 1}
\end{equation}
is a matrix of size $(2\ell + 1) \times (2\ell' + 1)$ indexed by $m_1 = -\ell \ldots \ell$ and $m_2 = -\ell' \ldots \ell'$, and $(\mathcal{A}_\ell)_{m, r} = A_{\ell, m}(r)$ is a matrix of spherical harmonics coefficients indexed by $m, r$. Here, the sum ranges are detailed in Eq.~\eqref{eq:gnarly_summation_ranges}. Since $\mathcal{B}$ is linear in $B$, we use \Cref{alg:BLSalg} to construct many linear systems $\mathcal{U}^n_{\ell, \ell'}$ such that: 
$$\mathcal{U}^{n}_{\ell, \ell'} \text{vec}(B)  = \text{vec}\left(\mathcal{B}^{n}_{\ell, \ell'}\right),$$
Using the Kronecker product, Eq.~\eqref{eq:m2_compact} can be written as
$$ \text{vec}\left(\mathbf{m}_2(\Delta \phi)\right) = \sum_{n} e^{in\Delta \phi} \sum_{\ell, \ell'} (\mathcal{A}_{\ell'} \otimes \mathcal{A}_{\ell}) \mathcal{U}^{n}_{\ell, \ell'} \text{vec}(B). $$
This, too, is linear in $B$:
\begin{equation} \label{eq:Umatrix}
\text{vec}\left(\mathbf{m}_2(\Delta \phi)\right) = U(\Delta \phi) \text{vec}(B), \quad \text{where} \quad  U(\Delta \phi) = \sum_{n} e^{in\Delta \phi} \sum_{\ell, \ell'} (\mathcal{A}_{\ell'} \otimes \mathcal{A}_{\ell}) \mathcal{U}^{n}_{\ell, \ell'}. \end{equation}
By vertically appending $V$ and copies of $U(\Delta \phi)$ for all values of $\Delta \phi$ in~\Cref{sec:volume-metric}, we obtain the least-squares formulation

$$ \min_x \|Ax - b\|,
$$
where \begin{equation}\label{eq:definedLSMatrix}
    A = \begin{pmatrix} V \\ U(\Delta\phi_1) \\ \vdots \\ U(\Delta\phi_n) \end{pmatrix}, \quad x = \text{vec}(B), \quad \text{ and } b = \begin{pmatrix} \text{vec}(\mathbf{m}_1) \\ \text{vec}(\mathbf{m}_2) \end{pmatrix}.
\end{equation}
To solve this, we perform $QR$-decomposition $A = QR$, and then solve the normal equations
$$A^*Ax = R^* Rx = A^*b = R^* Q^* b,$$ i.e., we solve $Rx = Q^*b$. Since $R$ is a square upper triangular matrix, we solve this using back substitution.

\subsection{Change of bases for moment comparison}\label{app:basis} 
We compute moments from images using the fast method\cite{fastpca} that produces the moments expanded in the Fourier Bessel basis. 
Thus, a change of bases is required for moment comparison. The Fourier Bessel basis has several nice properties that make it advantageous to use when computing the moment from images; it is orthonormal, frequency-ordered, steerable, provides fast radial convolutions, and has a fast transform\cite{fastfourierbessel}. The Fourier Bessel basis functions can be written in polar coordinates $(r,\theta)$ as 
\begin{equation}
 \psi_{n, k}(r, \theta) =   c_{n,k} J_n(\lambda_{nk} r) e^{in\theta},
\end{equation}
where $J_n$ is a Bessel function of the first kind of order $n$, and $\lambda_{nk}$ is the $k$-h smallest positive zero of $J_n$, and $c_{n,k}$ is a normalization constant. 

We create a change of basis matrix $(B)_{(r, \theta), (n, k)}= \psi_{n, k}(r, \theta)$ by sampling on a Cartesian grid $(x, y) \in \{r_i\} \times \{r_i\}$ with the $\{r_i\}$ grid defined as in \Cref{sec:volume-metric}, where $(r, \theta)$ are the grid points $(x, y)$ in polar coordinates. This yields the moments in real space \begin{align*}
    \mathbf{m}_1(\Phi, \rho)(x, y) &= B \mathbf{m}_1(\Phi, \rho)(n, k), \\
    \mathbf{m}_2(\Phi, \rho)(x_1, y_1, x_2, y_2) &= B \mathbf{m}_2(\Phi, \rho)(n_1, k_1, n_2, k_2) B^* .
\end{align*}
Now we compute the NUFFT to convert the moments into radially sampled polar coordinates in Fourier space as in Eq.~\eqref{eq:discrete_fourier_spherical}. In practice, we do this for $\mathbf{m}_2$ by taking each row (which is indexed by $(x_1, y_1)$), reshaping it into an image, and applying the transform. We then apply the same process to the columns indexed by $(x_2, y_2)$.

\subsection{Computational complexity} \label{app:complexity}
In the following, $L$ is the molecule bandlimit, see Eq. \eqref{eq:hatPhi}; $P$ is the distribution bandlimit, see Eq. \eqref{eq:rho}; $M$ is the number of projection images; and $N$ is the image side length of the $N \times N$ pixel images. We assume that $P \le L$.
 
There are three main steps for calculating the least squares matrix for each structure in our database. We first calculate the least squares matrices $\mathcal{U}^n_{\ell, \ell'}$ for $\mathcal{B}$ as described in \Cref{alg:BLSalg}. This needs only to be done once and does not need to be recomputed for each molecule. Calculating this matrix takes $O(P L^5)$ time and uses $O(P L^5)$ space. 
For the calculation of the least squares matrix itself, we precompute $ (\mathcal{A}_{\ell'} \otimes \mathcal{A}_{\ell}) \mathcal{U}^{n}_{\ell, \ell'}, $ for $\ell, \ell', n$ as described in Eq.~\eqref{eq:Umatrix}. These intermediate steps take $O(P^2 L^2 N^2 + L^3 N^2)$ time and use $O(L^3 N^2)$ space for forming the Kronecker product and subsequent matrix multiplication. Finally, the construction of the least squares matrix $A$ in Eq.~\eqref{eq:definedLSMatrix} takes $O(L^3 N^3)$ time for the scalar multiplication of a matrix for each $n, \ell, \ell'$, and the least squares uses $O( P^2 N^3)$ space. As such, the total computational complexity for calculating a least-squares matrix is $O(P^2L^5 + P^2L^2N^2 + L^3N^3)$ time and $O(P^2N^3 + L^3N^2 + PL^5)$ space.
 
The computation of moments from noisy projection images in the Fourier-Bessel basis takes $O(MN^3 + N^4)$ time and uses $O(MN^2 + N^3)$ space. To convert this to polar coordinates in Fourier space, we must first evaluate the moments in the Fourier-Bessel basis. This takes $O(N^2 \log N)$ time for each expansion, and we require $2N^2$ such expansions (see \Cref{app:basis}). Hence, in total, this step takes $O(N^4 \log N)$ time, and uses $O(MN^2 + N^4)$ space. Converting into Fourier space using the NUFFT takes $O(N^4\log N)$ time and uses $O(N^4 \log N)$ space, and we do this $2N^2$ times for a total of $O(N^6 \log N)$ time and space complexity. Storing the final moment uses $O(N^3)$ space (since the resulting matrix from the NUFFT is block circulant). Overall, computing moments from images and converting them to polar coordinates in Fourier space takes $O(N^6 \log N + MN^3)$ time and uses $O(MN^2 + N^6 \log N)$ space. 

\subsection{NDCG Score}\label{app:NDCG}

The NDCG \cite{ndcg} is calculated by taking the Discounted Cumulative Gain (DCG) and normalizing by Ideal Discounted Cumulative Gain (IDCG):
\begin{equation}
\begin{split}
    DCG &= \sum_{i} \frac{\text{true score of item $i$}}{\log(i + 1)}, \\
    IDCG &= \sum_{i} \frac{\text{$i^{th}$ highest true score}}{\log(i + 1)}, \\
    NDCG &= \frac{DCG}{IDCG},
\end{split}
\end{equation} where $i$ is enumerated in the order induced by the predicted scores.

The NDCG puts weight on scores that are agreed to be high by both metrics. However, our metric and the metrics we compare to are dissimilarity scores, so we prefer weight on scores that are considered low by both metrics. To remedy this, we use the reverse of the order enumerated by the predicted scores. For the true scores, we first normalize the scores to the range $[0, 1]$ and then take the exponential $e^{-s}$ for each true score $s$.

\section{Additional Results}
\subsection{Parameter selection}
In the experiments, we set the bandlimit parameters to $P = 6$ and $L = 25$. Note that this value of $P$ is comparable to previous work as described in \cite{nonuniform}, whereas the higher value of $L$ allows for a more accurate representation of the molecule in spherical harmonics. Furthermore, the hyperparameter $\lambda$ was set to be 1. As shown in~\Cref{tab:lambda} below, varying $\lambda$ does not greatly impact the performance of the metric.
\label{app:hyperparameter}
\begin{table}[!h]
\centering
\caption{Effect of the value of the hyperparameter $\lambda$ on the ranking induced by $d_{\operatorname{iKam}}$.
        $A_i$ denotes the structure with the $i$th lowest value of $d_{\operatorname{iKam}}$. In each row, the entry shaded green indicates the ground truth structure.} \label{tab:new_table}
        \label{1L1}
        \begin{tabular}{c ||c c c c c}
            $\lambda$ & $A_1$ & $A_2$ & $A_3$ & $A_4$ & $A_5$ \\ [0.5ex] 
            \hline\hline
            1e-2&  \cellcolor{green!25}\textbf{7VV3} & 7VUZ & 7TRK & 7TRP & 7VDM\\
            \hline
            1e-1&  \cellcolor{green!25}\textbf{7VV3} & 7VUZ & 7TRK & 7TRP & 7VDM\\
            \hline
            1&  \cellcolor{green!25}\textbf{7VV3} & 7VUZ & 7TRK & 7TRP & 7VDM\\
            \hline
            1e1&  \cellcolor{green!25}\textbf{7VV3} & 7VUZ & 7TRK & 7TRP & 7VDM\\
            \hline
            1e2&  \cellcolor{green!25}\textbf{7VV3}&   7VUZ&   7TRK&  7TRP&   7VDM\\
            \hline
            1e3&  7Y15 & 6K41 & 7VDM & 7VUZ & 7E33 \\
\end{tabular}
\label{tab:lambda}
\end{table}

\subsection{Additional t-SNE plots}\label{app:tsne}
This appendix includes t-SNE visualizations of the Zernike metric on our database and the alignment metric on a subset of our database of size 100. The alignment metric is restricted to a subset of size 100 since calculating pairwise distances for a 1420 is computationally taxing, see~\Cref{sec:database}. Visually, the Zernike t-SNE seems to have fewer distinct clusters than the t-SNE plot generated using $d_{\operatorname{vKam}}$, and also groups molecules with different numbers of atoms together. It seems possible that Zernike metric is less discriminative, although this may also be an artifact of t-SNE's dimensionality reduction.
\begin{figure}[!h]
	\includegraphics[width= 1 \textwidth]{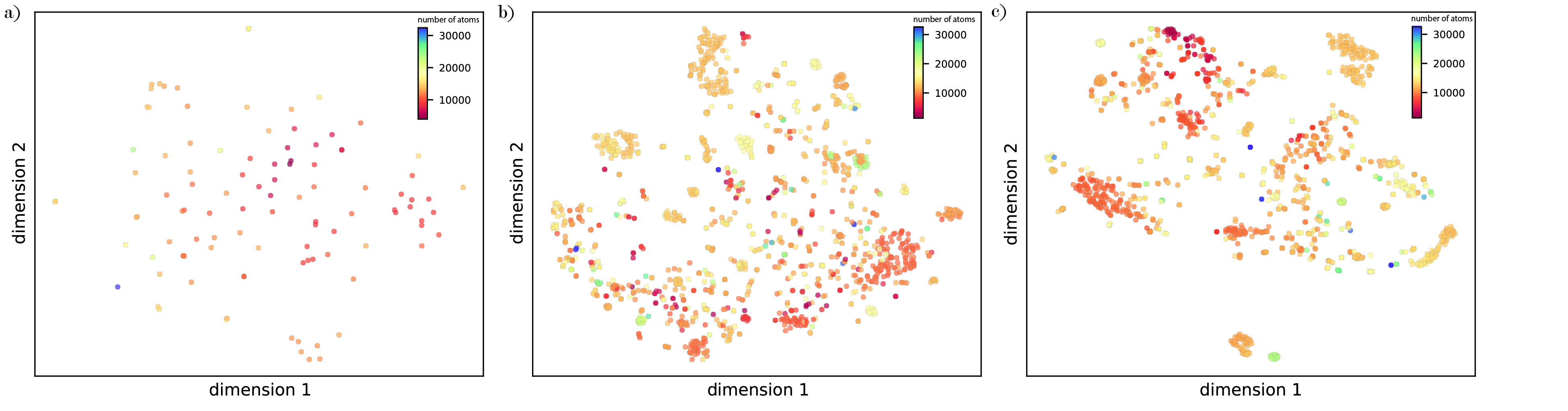}
	\centering
	\caption{Additional t-SNE plots.
 a) t-SNE plot of pairwise Euclidean alignment distances on a subset of size 100. b) t-SNE plot of pairwise Zernike distances on the entire database c) t-SNE plot of pairwise $d_{\operatorname{vKam}}$ distances on the entire database}
\end{figure}
\subsection{Robustness to number of images} \label{app:robustness}
 Here we examine the robustness of our metric to inaccuracies of moment estimation. Specifically, we vary the number of noisy synthetic projection images that the metric has access to and record the highest-ranking structures.
\begin{table}[!h]
\centering
\caption{ Effect of the number of projection images used for moment estimation on the ranking induced by $d_{\operatorname{iKam}}$.
        $A_i$ denotes the structure with the $i$th lowest value of $d_{\operatorname{iKam}}$. In each row, the entry shaded green indicates the ground truth structure. The relative error in each moment is between the moment estimated from noisy projection images and the moment calculated from their clean counterparts.}
\resizebox{\textwidth}{!}{
        \label{1L1}
        \begin{tabular}{c ||c c c c c ||c c}
            number of images & $A_1$ & $A_2$ & $A_3$ & $A_4$ & $A_5$ & relative $M_1$ error (\%) & relative $M_2$ error (\%)\\ [0.5ex] 
            \hline\hline
            500&  7VDM&   7VUZ&   7Y15&  7E33&   \cellcolor{green!25}\textbf{7VV3}&   1.49& 8.23\\
            \hline
            1000&  7VUZ&   7VDM&   \cellcolor{green!25}\textbf{7VV3}&  7TRK&   7EJ8&   0.76& 6.22\\
            \hline
            2500&  7VUZ&   7VDM&   \cellcolor{green!25}\textbf{7VV3}&  7TRK&   7TRP&   0.43& 4.02\\
            \hline
            5000&  \cellcolor{green!25}\textbf{7VV3}&   7TRK&   7VUZ&  7TRP&   7VDM&   0.27& 2.95\\
            \hline
            10000&  \cellcolor{green!25}\textbf{7VV3}&   7VUZ&   7TRK&  7TRP&   7VDM&   0.25& 1.89\\
            \hline
            25000&  \cellcolor{green!25} \textbf{7VV3}&   7VUZ&   7TRK&  7VDM&   7TRP&   0.19& 1.37\\
            \hline
            50000&  \cellcolor{green!25} \textbf{7VV3}&   7VUZ&   7TRK&  7VDM&   7TRP&   0.14& 1.15
        \end{tabular}}
        \label{tab:robustness-images}
\end{table}

\subsection{Additional experimental results} \label{app:experimental}
\Cref{tab:robustness-images} reports the metric's rankings using experimental images corresponding to the five structures resolved from EMPIAR-10076.
\begin{table}[!h]
\centering
\caption{Performance of $d_{\operatorname{iKam}}$ on structures 001, 002, 003, 004, 005 of EMPIAR-10076.
        $A_i$ denotes the structure with the $i$th lowest value of $d_{\operatorname{iKam}}$. The value of $\log(d_{\operatorname{iKam}})$ is reported next to each structure in parenthesis. In each row, the entry shaded green indicates the ground truth structure.}
\resizebox{\textwidth}{!}{
        \label{1L1}
        \begin{tabular}{c ||c c c c c c c}
             number of images & $A_1$ & $A_2$ & $A_3$ & $A_4$ & $A_5$ & $A_6$ & $A_7$ \\ [0.5ex] 
            \hline\hline
            2018 & \cellcolor{green!25}\textbf{000} (14.56) & 004 (16.22) & EMD-2660 (16.23) & 003 (16.32) & 002 (16.42) & 001 (16.45) & EMD-8457 (17.43) \\
            \hline
            12650 & \cellcolor{green!25}\textbf{001} (14.28) & 002 (14.29) & 004 (14.34) & 003 (14.35) & 000 (14.95) & EMD-8457 (15.15) & EMD-2660 (15.64) \\
            \hline
            26104 & 001 (15.13) & \cellcolor{green!25}\textbf{002} (15.16) & 004 (15.18) & 003 (15.29) & 000 (16.50) & EMD-8457 (16.51) & EMD-2660 (17.14) \\
            \hline
            26138 & \cellcolor{green!25}\textbf{003} (14.56) & 004 (14.73) & 001 (14.74) & 002 (14.74) &   000 (15.50) & EMD-8457 (15.83) & EMD-2660 (16.08) \\
            \hline
            36561 & 003 (14.62) &  \cellcolor{green!25}\textbf{004} (14.80) & 001 (14.84) & 002 (14.88) & 000 (15.40) & EMD-8457 (16.00) & EMD-2660 (16.09) \\
            \hline
        \end{tabular}}
        \label{tab:robustness-images}
\end{table}

\end{document}